\documentclass[a4paper,12pt]{article}
\usepackage{epsfig}
\usepackage{graphicx}
\usepackage{subfigure}
\begin{document}
\title{\textbf{Scattering of Topological Solitons on Barriers and Holes of Deformed Sine-Gordon Models}}
\author{Jassem H. Al-Alawi\thanks{e-mail address:J.H.Al-Alawi@durham.ac.uk} and Wojtek J. Zakrzewski\thanks{email address: W.J.Zakrzewski@durham.ac.uk}
\\ Department of Mathematical Sciences,University of Durham, \\
 Durham DH1 3LE, UK\\
}

\date{\today}


\maketitle

\begin{abstract}
We study scattering properties of topological solitons in two classes of models, 
which are the generalisations of the Sine-Gordon model and which have recently
been proposed by Bazeia et al. These two classes of models
depend on an integer parameter $n$ which, when $n=2$ (for the first class)
and $n=1$ (for the second class), reduce to the Sine-Gordon model.

We take the soliton solutions of these models (generalisations of the `kink'
solution of the Sine-Gordon model) and consider their scattering
on potential holes and barriers. We present our results for $n=1,...6$.

We find that, like in the Sine-Gordon models, the scattering on the barrier is very elastic while the scattering on the hole is inelastic and can, at times, 
lead to a reflection. 

We discuss the dependence of our results on $n$ and find that the critical velocity 
for the transmission through the hole is  lowest for $n=3$.

\end{abstract}

\section{Introduction}
 Solitons, as solutions of non-linear wave equations have, by now, been studied extensively for many years \cite{one}. However, they are still producing
  new and unexpected phenomena. Such is, for instance, their behaviour when
 one sends a soliton towards a potential obstruction  as was studied in \cite{two}, \cite{three}. The
 results presented in these papers were obtained for solitons in (2+1) dimensions and for the Sine-Gordon model
 in (1+1) dimensions. 
 
 In a recent study \cite{four} we looked at a similar scattering of solitons
  in two (1+1) dimensional $\varphi^{4}$ models. In these models we have inserted 
  potential obstructions in two different ways, in each case
  modifying the Lagrangian of the model in a region far away from the soliton. In the first model this modification was introduced via the coupling of the potential
  (which was made to be position dependent), while in the second model this
   was achieved via the modification of the Minkowski space-time metric \cite{five}.
   
  In each model, like in \cite{two} and \cite{three}, the topological solitons 
  were found to scatter on the barrier in a very elastic way: they can either overcome the barrier and get transmitted or get reflected from the barrier with almost no loss of energy. 
  However, the scattering from a hole was found to be inelastic and, at times,
  produced an almost non-classical behaviour as solitons with energies below a critical value for the transmission could either be trapped in the hole or get reflected!  
  This reflection resembles a little the quantum reflection and so, perhaps,  some classical phenomena based on solitons could be confused with their quantum behaviour.

In view of this we have decided to look at other models; to see whether
the observed phenomena are `universal'. A good class of such models is provided 
by the models of Bazeia et al \cite{six}. Not only they depend on a parameter
(which takes integer values); they also reduce to the well known Sine-Gordon
model for a specific value of this parameter. Hence in this paper we look
at the scattering of solitons in these models on both potential holes and potential
barriers. Like in the original work on the Sine-Gordon model \cite{three}
the obstructions are introduced via the modification of the coupling
constant in the models {\it i.e.} by making it space dependent.

In the next section we introduce the models of Bazeia et al \cite{six}
and discuss some of their properties. We present the relevant Lagrangians
(for both classes of generalizations of the Sine-Gordon model) 
and discuss their soliton (kink-like) solutions.
The following two sections discuss our results on the scattering
of these solitons (from each class) on both types of obstructions.
We finish the paper with a short section presenting our conclusions.

\section{\textbf{The models of Bazeia et al}}
We consider, in full generality,  a single real scalar field in (1+1) dimensions

\begin{equation}
\ell=\frac{1}{2}\partial_{\mu}\varphi\partial^{\mu}\varphi-\,\tilde V\left(\varphi\right),
\label{lag}
\end{equation}
where the potential $\,\tilde V\left(\varphi\right)$ is taken in the form
\begin{center}
 $\,\tilde V\left(\varphi\right)=\tilde \lambda^{2}\,V\left(\varphi\right)$,
\end{center}
in which  $\tilde \lambda=1+\lambda\left(\,x\right)$.
 
\noindent Here $\lambda\left(\,x\right)$ is an extra potential parameter which has been inserted into the potential $\,\tilde V\left(\varphi\right)$ to take into account the effects of  obstructions, holes and barriers, and so is nonzero only in a certain region of space.

In our case, we  put the obstruction around the origin ({\it i.e.} $x=0$)
so we  take
\begin{center}
$\lambda\left(\,x\right)=\cases{ 0 & $\vert\,x\vert> 5 $ \cr
                                 \lambda_{0}=\hbox{constant} & $\vert\,x\vert\leq 5. $ \cr} $
\end{center}

The equation of motion is, of course,
\begin{equation}
 \partial_{\mu}\partial^{\mu}\varphi + \,\tilde {V}'\left(\varphi\right)=0,
\end{equation}
where $\,\tilde {V}'$ is the derivative of $\,\tilde V$ with respect to the argument.

The models of Bazeia et al \cite{one} correspond to choosing
(`type 1' models) $\tilde V$ of the form:
\begin{equation}
 \,\tilde V\left(\varphi\right)=\frac{2\tilde \lambda^{2}}{\,n^{2}}\tan^{2}\left(\varphi\right)\left(1-\,sin^{\,n}\left(\varphi\right)\right)^{2}.
\end{equation}
 When $n=2$ this model reduces to the Sine-Gordon one.

The second class (`type II models') correspond to taking $\tilde V$ in the form
\begin{equation}
 \,\tilde V\left(\varphi\right)=\frac{\tilde \lambda^{2}}{2\,n^{2}}\varphi^{2-2\,n}\,sin^{2}\left(\varphi^{n}\right).
\end{equation}

In this case the Sine-Gordon model corresponds to  $n=1$.

Both classes of models are topological as in each case we can write for the static fields:
\begin{equation}
\ell=-\frac{1}{2}\partial_{x}\varphi\partial_{x}\varphi-\,\tilde V\left(\varphi\right)\,=\,
-\frac{1}{2}\left( \partial_x\varphi\pm \sqrt{2}\sqrt{\tilde V}\right)^2\,\pm\,\frac{1}{\sqrt{2}}
\partial_x\varphi\sqrt{\tilde V}.
\end{equation}

The last term is a total divergence and so we see that the equation for the static
solitons (Bogomolny'i equation) becomes
\begin{equation}
\partial_{x}\varphi=\,\pm \sqrt{2}\sqrt{\tilde V\left(\varphi\right)} .
\label{bogo}
\end{equation}


 

The solutions of (\ref{bogo}) are easy to find in the case when we have no
obstruction ({\it i.e.} when $\tilde \lambda=1$). They are given
by (for the first class of models):
\begin{equation}
 \varphi\left(\,x,\,t\right)=\,sin^{-1}\left[\frac{\,exp\left(2\tilde \lambda\gamma\left(\,x-x_{0}-\,u\,t\right)\right)}{1+\exp\left(2\tilde \lambda \gamma\left(\,x-x_{0}-\,u\,t\right)\right)}\right]^\frac{1}{\,n},
 \label{onee}
\end{equation}
where, due to the Lorentz invariance of the full Lagrangian (\ref{lag}),
we have inserted the time dependence by performing the relativistic boost.
In (\ref{onee}) $\gamma$ is the usual relativistic factor, {\it ie} $\frac{1}{\sqrt{1-\,u^{2}}}$.
The expression (\ref{onee}) is, in fact, a solution of (\ref{bogo}) when $\tilde \lambda=1$. It is
also an approximate solution of (\ref{bogo}) when $\tilde \lambda\ne1$ for $x_0$ far away from the
region in which $\tilde\lambda\ne1$.

The soliton solutions (\ref{onee}) satisfy the kink boundary conditions
\begin{center}
 $ \varphi\left(\,x\rightarrow-\infty\right)\longrightarrow0, \quad \varphi\left(\,x\rightarrow\infty\right)\longrightarrow\frac{\pi}{2}$.
\end{center}

For the second class of models the soliton solutions (again with the same comments about
the $x$ dependence of $\tilde \lambda$) are given by:

\begin{equation}
 \varphi\left(\,x,\,t\right)=\left[2\,tan^{-1}\left(\,exp\left(\tilde \lambda \gamma\left(\,x-\,x_{0}-\,u\,t\right)\right)\right)\right]^\frac{1}{\,n},
\end{equation}
and they satisfy  the following boundary conditions:
\begin{center}
 $ \varphi\left(\,x\rightarrow-\infty\right)\longrightarrow0, \quad \varphi\left(\,x\rightarrow\infty\right)\longrightarrow\pi^\frac{1}{n}$.
\end{center}

In figures 1 and 2 we present plots of the static soliton field configurations for the type I and type II models, respectively, for the first 6 models in each class, {\it i.e.} for  $n=1,...,6$.

In the next section we look at the scattering properties of these solitonic solutions
in the first class of models.

\begin{figure}
\begin{center}
\includegraphics[angle=270, width=8cm]{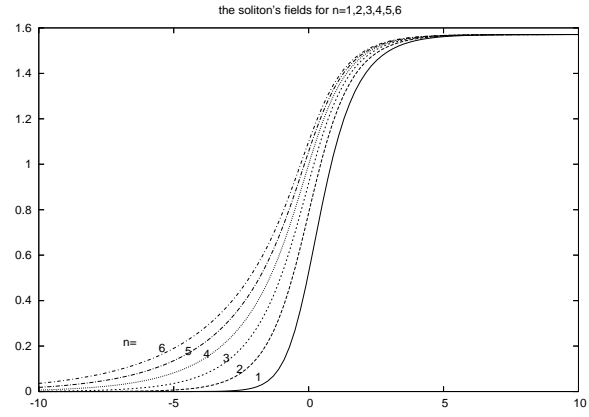}
\caption{Type-I models: Soliton fields for $n$=1,...,6.}
\end{center}
\end{figure}

\begin{figure}
\begin{center}
\includegraphics[angle=270, width=8cm]{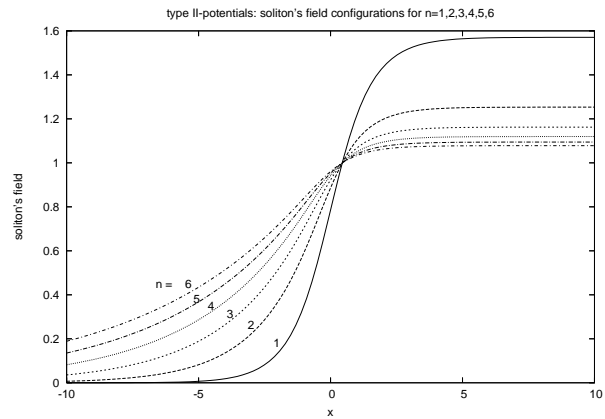}
\caption{Type-II models: Soliton fields for $n$=1,...,6.}
\end{center}
\end{figure}

\section{\textbf{Solitons Scattering of Type-I Potentials}}  
Before we consider the scattering properties of the solitonic solutions let us observe that the mass
(rest energy) of a static soliton is given by
\begin{equation}
\,M_{rest}=\int_{-\infty}^{\infty}\left[\frac{1}{2}\varphi_{x}^{2}+\,\tilde V\left(\varphi\right)\right]\,dx.
\end{equation}  
The energy of a soliton moving with velocity $u$  is then given by
 
\begin{equation}
 \,E=\frac{\,M_{rest}}{\sqrt{1-\,u^{2}}}.
\end{equation}

To calculate $M_{rest}$ we note that it can be rewritten as
\begin{equation}
\,M_{rest\left(n\right)}=\frac{4}{\,n^{2}}\int_{-\infty}^{\infty}\frac{\,W^{\frac{2}{\,n}}\left(1-\,W\right)^{2}}{\left(1-\,W^{\frac{2}{\,n}}\right)}\,dx,
\label{mass}
\end{equation}
where $W$ is
\begin{center}
$ \,W=\frac{\,exp\left(2\,x\right)}{1+\,exp\left(2\,x\right)}$.
\end{center}

To perform the calculation of (\ref{mass}) it is convenient to change variables
from $x$ to $r=W^{\frac{1}{n}}$.
Then (\ref{mass}) becomes proportional to $S_n$ {\it i.e.}
\begin{equation}
M_{rest}=2\tilde\lambda^2 S_n,\quad\hbox{where}\quad S_n\,=\,\frac{1}{n}\int_0^1 \frac {r(1-r^n)}{1-r^2}dr.
\end{equation}
 
This integral defining $S_n$ can be performed explicitly for each value of $n$. We find that
\begin{equation}
S_1\,=\,1-\ln 2,\qquad S_2\,=\,\frac{1}{4}
\end{equation}
and all others satisfy the recurrence relation
\begin{equation}
S_{n+2}\,=\,\left[n S_{n}+\frac{1}{n+2}\right]\frac{1}{n+2}.
\end{equation}

In figure 3 we plot the rest mass energies for $n=1,...,6$. The figure shows that  solitons becomes less massive as $n$ increases. The figure also shows that the $n=1$ the soliton is very massive.

\begin{figure}
\begin{center}
\includegraphics[angle=270, width=8cm]{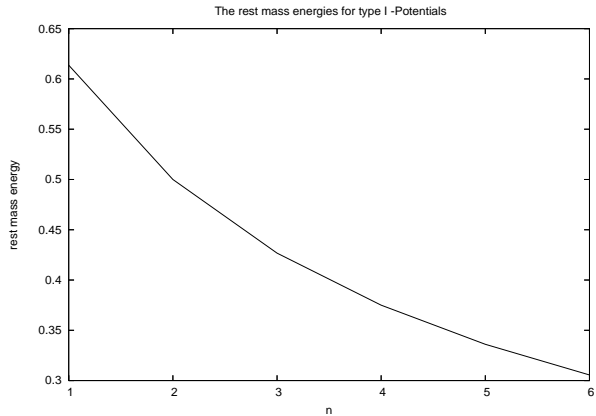}
\caption{Type-I potentials: Rest mass energies for $n=1,...,6.$}
\end{center}
\end{figure}

Next we considered the scattering properties of these solitons. To do this we put our obstructions
close to $x=0$ (in fact between -5 and 5) and initially placed solitons far away from $x=0$, namely
around $-40$ sending them with some velocity $u$ towards the obstruction. In our numerical work  we used the square well potential barriers and holes of width 10. The width was chosen carefully so that the solitons can fit into the potential
hole and have enough space to move inside the hole or on the top of the barrier. The simulations were performed using the 4th order Runge - Kutta method of simulating the time evolution. We used 1201 points with the lattice spacing of  $dx=0.01$. Hence, the lattice extended from -60 to 60 in the $x$-direction. The time step was chosen to be $dt=0.0025$. In our work we used the absorbing boundary conditions.

\subsection{Hole Scattering}

Our simulations have shown that, like in the previous work on the Sine-Gordon \cite{three}
and $\lambda\phi^4$ \cite{four} models, for any $n$, there is a critical velocity
above which the solitons are transmitted by the hole and below which they are either
trapped or reflected. See also \cite{older} and \cite{Goodman} which present earlier work on this subject
with an interesting explanation of the reflection property of the hole
in terms of the interference of the soliton field with the radiation waves
generated by the scattering.  
 
 Hence we believe that the observed phenomenon is a generic behaviour of topological solitons as they encounter a potential hole.
The value of the critical velocity depends on the model
 in question. The scattering is inelastic with the soliton emitting radiation both
 in the hole and/if  when it has left the hole.
 
The amount of the emitted radiation is expected to be related to the integrability, or not, of the
model in question. Hence to study this further we have decided to look in detail
on the emission of radiation in the models of Bazeia et al.
Thus we have examined in detail the scattering properties of solitons for $n=1,..6$.
 
First we looked at the values of the critical velocity for the hole
 of depth =-0.50 ({\it i.e.} $\lambda_0=-0.5$). Figure 4 gives the plot of these values  and we note
 that  the critical velocity is the lowest for $n=3$.

In order to have a better understanding to our results on the critical velocities, we have looked into the energy of solitons after the scattering when they are already far away from the hole. The total energy of the  solitons after the scattering is given by

\begin{equation}
 \,E_{final}=\frac{\,M_{rest}}{\sqrt{1-\,u_{out}^{2}}}+\,E_{vibration}+\,E_{radiation}
\end{equation}
  
 We have estimated the ratios of the radiation to the vibrational energies. Table 1 presents
the results of our calculations.

\begin{center}
\begin{tabular}{ll}
n & $\,E_{radiation}/\,E_{vibration}$ \\
1 &  0.76050 \\
2 &  0.07069\\
3 &  0.05382\\
4 &  0.07711\\
5 &  0.08120 \\
6 &  0.09752
\end{tabular}
\end{center}

The table clearly does not exibit any regular pattern for this ratio which would have encouraged us
to seek a deeper understanding of the mechanism of this process (in terms of possible
equipartition of energy etc). Unfortunately, the results of the table suggest that any
an explanation will be difficult to find. We do note, however, that in the $n=3$ this ratio is the
smallest. This shows that in this case the solution preserves more  of its extra energy
as a vibrational energy and so has it available when the soliton tries
to come out of the hole. Hence the critical velocity is also
the lowest for $n=3$.

\begin{figure}
\begin{center}
\includegraphics[angle=270, width=8cm]{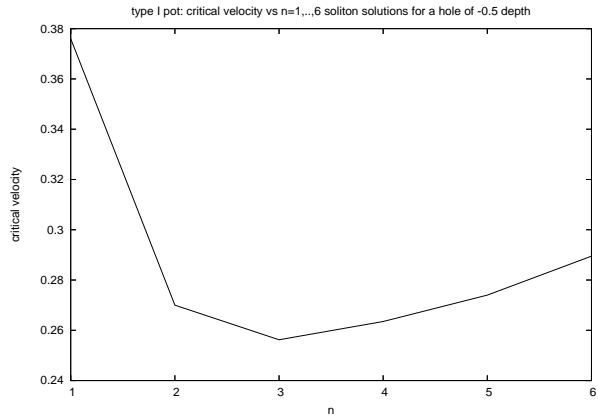}
\caption{Type-I models: Critical velocities of soliton solutions for a hole of -0.5 depth; first six models.}
\end{center}
\end{figure}

We have also looked at the behaviour of solitons as they approach this hole (of -0.5 depth) with velocities above the critical values, {\it i.e.} when they get transmitted. Figure 5
presents the plots of the time dependence of the positions of these solitons for $u$=0.5 and $u$=0.7. One sees very clearly that for velocities well above
its critical value all solitons behave almost in the same way. In each case the
velocity of the transmitted soliton is almost the same (but, of course, always lower
that the initial velocity). The only exception is the $n=1$ case when the transmitted
soliton has considerably lower velocity - but then for $n=1$ $u=0.5$ is not that
far away from the critical velocity for this model. Hence we conclude that for velocities significantly
above its critical value the velocity of the transmitted soliton does not
depend much on $n$.

\begin{figure}
\begin{center}
\includegraphics[angle=270, width=8cm]{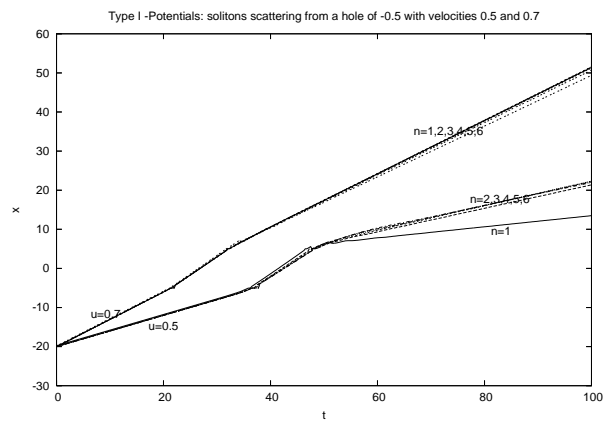}
\caption{Type I - models: Trajectories of solitons for velocities
above their critical values; {\it ie} $u=0.5$ and $u=0.7$ for a hole of depth -0.5.}
\end{center}
\end{figure}

We have also looked at the energy loss of the transmitted soliton. In this case some
energy is radiated away when the soliton is in the hole.
When it emerges it is excited - {\it i.e.} some of its energy
is converted into the excitation energy which is then slowly radiated away
as the soliton moves away from the hole.

To confirm these expectations we have looked at the velocity of the soliton after it has left the hole
and at its total energy.
Thus, for example,  looking  at the soliton of the $n=3$ model moving towards the
hole we note that its total energy is given by:

\begin{equation}
\,E_{in\left(3\right)}=\frac{\,M_{rest\left(3\right)}}{\sqrt{1- \,u_{in}^{2}}}=\frac{0.4276888}{\sqrt{1-0.5^{2}}}=0.4928.
\end{equation}
 
After the scattering the observed  velocity of the soliton is 0.43999. For such a velocity (assuming that there is no excitation) the total energy would be
\begin{equation}
\,E_{out\left(3\right)}=\frac{\,M_{rest\left(3\right)}}{\sqrt{1- \,u_{out}^{2}}}=\frac{0.4276888}{\sqrt{1-0.43999^{2}}}=0.476266.
\end{equation}

However, the observed energy of the system, after the soliton
has left the hole, was pretty much the same as the initial energy, {\it i.e.} 0.4928.
 The difference between these two energies, {\it i.e.} 0.01653, must be the excitation energy of the soliton and of the radiation that has been emitted
 during the scattering process, which has still not been able
 to reach the boundaries of our grid (we absorb at the boundaries).
 In fact, the soliton radiates away this excess of energy all the time so that only asymptotically
 its energy will drop to the value corresponding to its velocity, {\it i.e.} 0.476266.
  
 However, this process is very slow and the soliton reaches the boundary before its
 energy exibits a significant drop. Hence to study this effect we have decided to look
 at a static soliton in the hole (which would be excited as its `profile' (determined
 by $\tilde \lambda$ )is not correct).

 We have looked at the case of a soliton for $n=1$.
 Its energy is 0.417.  When we have placed it inside a hole
 of depth =-0.4 we have found that its energy is

\begin{equation}
\,M_{rest\left(1\right)}=\frac{4}{\,n^{2}}\int_{-\infty}^{\infty}\frac{\,W^{2}\left(1-\,W\right)}{\left(1+\,W\right)}\,dx=0.368223,
\end{equation}
as this time  $W$ is given by
\begin{center}
 $\,W=exp\left(1.2\,x\right)/\left(1+exp\left(1.2\,x\right)\right).$
\end{center}

Thus this soliton inside the hole is excited and so it radiates away its
excess of energy. This process is quite slow as shown in fig 7 where we plot the time
 dependence of the energy of such a soliton. We note that although the energy has dropped quite a lot
 even at $t=1000$ it is still significantly above its asymptotic (expected) value
 {\it i.e.} 0.368223.

How does the total excitation  + radiation energy depend on $n$?
This we show in figure 8 which presents a plot of the $n$ dependence of the difference between the observed total energy (just after the soliton left the hole) and the energy of a moving soliton after the scattering for the case when the soliton's incoming velocity was 0.6, which is far away from their critical value, for a hole depth of -0.5.

\begin{figure}
\begin{center}
\includegraphics[angle=270, width=8cm]{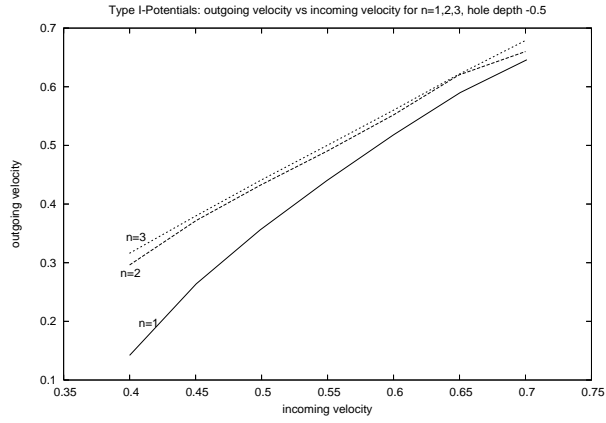}
\caption{Type-I models: ${\it v_{out}}$ vs ${\it v_{in}}$ for a hole of -0.5 for $n$=1,2,3.}
\end{center}
\end{figure}

\begin{figure}
\begin{center}
\includegraphics[angle=270, width=8cm]{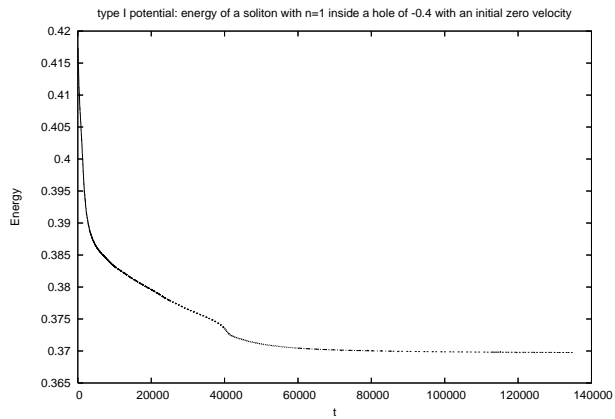}
\caption{Type-I models: The $n=1$ soliton's final energy inside a hole of -0.4, $u$=0.0.}
\end{center}
\end{figure}

\begin{figure}
\begin{center}
\includegraphics[angle=270, width=8cm]{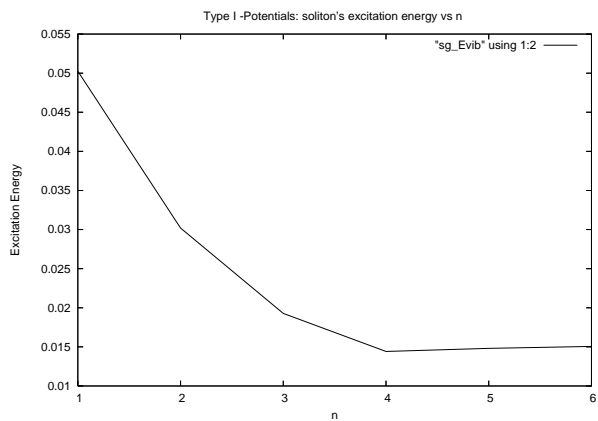}
\caption{Type-I models: Soliton excitation energy after the scattering from a hole of -0.5, $u$=0.6.}
\end{center}
\end{figure}

\subsection{Scattering on a Barrier}

The previous work \cite{two}, \cite{three} has found that the scattering
of topological solitons on potential barriers is very elastic ({\it i.e.}
very little energy was emitted and  the solitons behaved like point-particles).
Thus if the solitons had enough energy to `climb' to the top of the potential barrier
they were transmitted; otherwise they were reflected.

This is very much what we have seen in the models of Bazeia et al. When we looked at
the barrier of height 0.08 the outgoing and the incoming velocities of the solitons were essentially the same (within numerical errors). This was seen for all $n$ that we have looked at, {\it i.e.} $n=1,..6$.
 
Thus, we conclude that, like in all previous cases, the  scattering of topological solitons on barriers is almost totally elastic.


Next we calculated the critical values of velocity, as a function of $n$, for the scattering on a barrier of height 0.4. Our results are presented as a curve shown in
Figure 9.

\begin{figure}
\begin{center}
\includegraphics[angle=270, width=8cm]{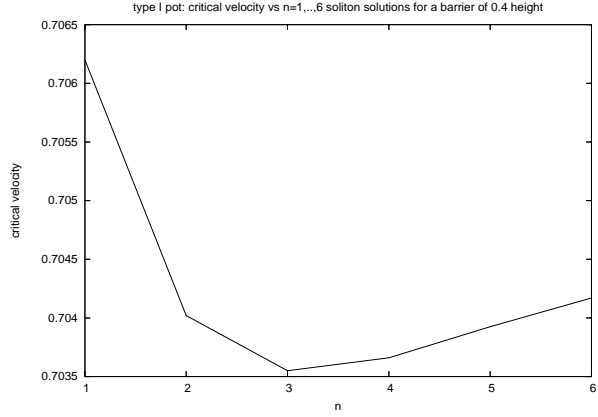}
\caption{Type-I -models:Critical velocities for the soliton solutions on a barrier of 0.4 height for the first 6 models.}
\end{center}
\end{figure}

Once again, {\it i.e.} for the hole, we see that the critical value is the lowest
for $n=3$. Hence the $n=3$ model is somewhat special - its critical velocities
are lowest both for the scattering on a hole and on a barrier.

As there is very little radiation and the scattering on barriers is very elastic the critical velocity can be estimated by approximating solitons by point particles.
 
The critical velocity, $v_{cr}$, is the velocity that a soliton has to have so that
it can climb the barrier; if the scattering is fully elastic this can be estimated
by looking at the energy of the soliton at rest at the top of the barrier $\,E_{cr\left(n\right)}$.

This energy is then approximately equal to the energy of a soliton away from the
barrier moving with velocity $v_{cr}$.

Thus
\begin{center}
$ \,E_{cr\left(n\right)}\sim \frac{\,M_{rest\left(n\right)}}{\sqrt{1-\,u_{cr\left(n\right)}^{2}}}.$
\end{center}
The critical energy is almost equal the rest mass energy of solitons at the top of a barrier. So,

\begin{center}
$ \,E_{cr\left(n\right)}\sim \,M_{B\left(n\right)},$
\end{center}
where $\,M_{B\left(n\right)}$ is the rest mass energy of a soliton at the top of the barrier. Thus the critical velocity is given by
\begin{equation}
\,u_{cr\left(n\right)}=\sqrt{1-\left(\frac{\,M_{rest\left(n\right)}}{\,E_{cr\left(n\right)}}\right)^{2}} .
\end{equation}

 In figure 10 we present the numerically calculated rest masses, for $n=1,..6$,  of the solitons  at the top of a barrier of 0.4 height.

\begin{figure}
\begin{center}
\includegraphics[angle=270, width=8cm]{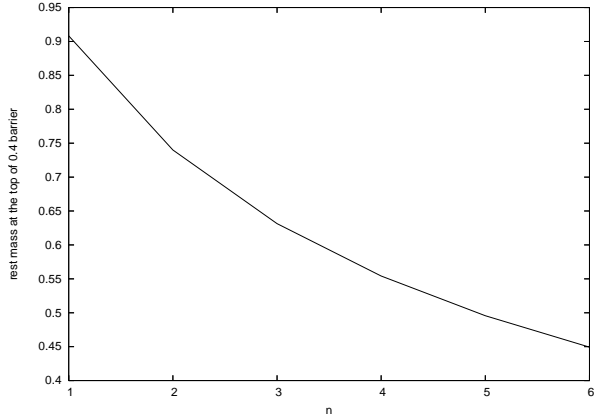}
\caption{Type-I models: $n$ dependence of the rest masses of  solitons for a barrier of  height 0.4.}
\end{center}
\end{figure}

Given these values we can calculate the approximate values
of the critical velocity and compare them with the actual values determined
numerically. The results are in close agreement.

Table 2 shows the calculated critical velocity for the soliton solutions
of the first six models ({\it i.e.} for $n=1,..,6$). We have not used the numerical values of the critical energies because solitons at the top of a barrier, with a zero velocity, are excited and so their energies are a marginally higher than their critical values.

\begin{center}
\begin{tabular}{llll}
n & $\,M_{rest}$ & $\,E_{cr}$ & $u_{cr}$ \\
1& 0.6137 & 0.8592& 0.699861 \\
2 & 0.5000 & 0.7000& 0.699854 \\
3 & 0.4268& 0.5975 & 0.699859\\
4 & 0.3749 & 0.525 & 0.699857 \\
5 & 0.3361 & 0.4705 & 0.699864\\
6 & 0.3056& 0.4278 & 0.699858
\end{tabular}
\end{center}

Clearly, both sets of values of critical velocities are in a very good agreement. At the same time, the
facts that the approximate value is lower, but only marginally so, and the value of this
difference, shows that the radiation effects are very small though, strictly speaking,
nonzero.

\section{\textbf{Soliton Scattering in the Type-II Potentials}}  

We have also looked at the scattering of solitons on  holes and barriers in the second
class of models of Bazeia et al. Most of the behaviour was very similar to what was seen
in the first class of models ({\it i.e.} the existence of critical velocities,
transmission and reflections on barriers etc).
But there were also some small differences. Hence here we restrict
our discussion to the description of these differences.

The main reason for these differences resides in the form of the soliton field itself, and the fact that, as shown in figure 2, the fields are very asymmetric with respect to their
behaviour as $x\rightarrow \pm \infty$. Clearly as $n$ increases the fields go to different, and decreasing values as $x\rightarrow \infty$. Similarly,
as $n$ increases, they go to 0 much more slowly. Hence the energy densities
of the solitons are more spread out, as shown in figure 11.

\begin{figure}
\begin{center}
\includegraphics[angle=270, width=8cm]{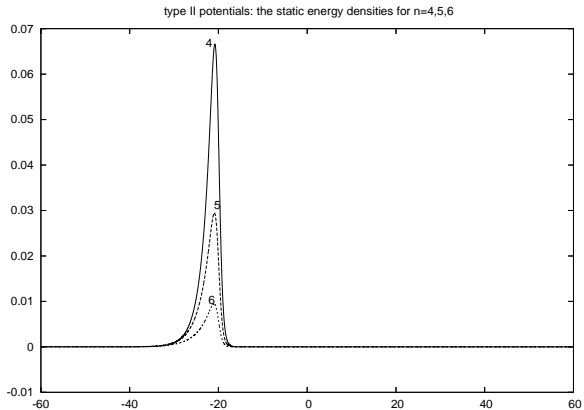}
\caption{Type-II potentials: Static energy densities for $n$=4,5,6.}
\end{center}
\end{figure}

Moreover, like for the first class of models, the rest masses of solitons decrease with $n$, but this time this decrease is faster. In Fig 12. we present the rest mass energies
of solitons as a function of $n$. Clearly, we see a big decrease as $n$ increases.

 As we will show below the tail of solitons in class II models, except for $n=1$ when the model reduces to the Sine-Gordon one,  affects  the behaviour of  solitons when they scatter on a potential hole.
  The effect increases with $n$ as for larger $n$ solitons are more spread out and also are less massive.
  

 In figure 11 we plot the energy densities of the  soliton solutions of Type II models. We only
 plot them for $n=4,5$ and $6$ as the energy densities for lower values on $n$ are much larger
 so that plotting them all together would make the plots of larger $n$ almost invisible.
 
 A note on the normalisation. As we said before the $n=2$ type I model  and $n=1$ type II one are the same
 - in fact they both correspond to the Sine-Gordon model. However, our values of their enegies are different
 (in the type I case the rest energy is 0.5, while in the type II it is 2.0). The difference
 comes from the different normalisation of both models ($\varphi\rightarrow 2\varphi$).

\begin{figure}
\begin{center}
\includegraphics[angle=270, width=8cm]{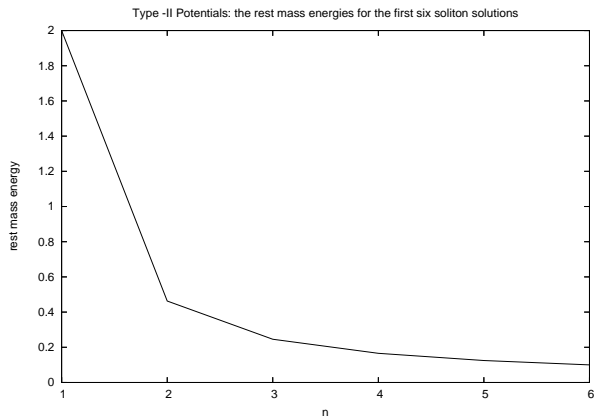}
\caption{Type-II potentials: Rest mass energies for $n$=1,...,6.}
\end{center}
\end{figure}

 Figures 13 and 14 present the plots of the critical velocities when the solitons are scattered from a hole of -0.5 depth and a barrier of 0.4 height, respectively.

\begin{figure}
\begin{center}
\includegraphics[angle=270, width=8cm]{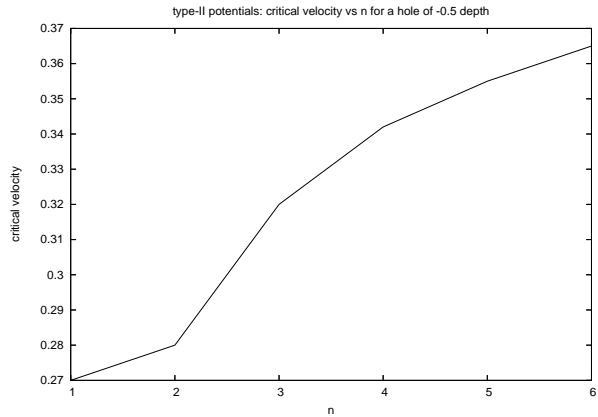}
\caption{Type-II models: Critical velocities for the soliton transmission through a hole of -0.5 depth for the first six models.}
\end{center}
\end{figure}

\begin{figure}
\begin{center}
\includegraphics[angle=270, width=8cm]{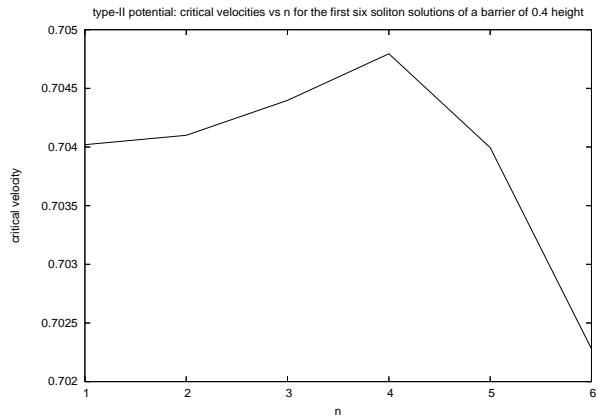}
\caption{Type-II models: Critical velocities for the soliton transmission over a barrier of 0.4 height
for the first six models.}
\end{center}
\end{figure}

\subsection{Tail Effect in Hole Scattering}

The tail, as we have observed in our numerical simulations, is the main part of the solitons  that is the most affected when the solitons are scattered on a hole. More precisely, our simulations have shown that the front part of the soliton is less affected by the hole while the tail, which is its rear part, is more affected by it. This is due to the asymmetry of the soliton field configuration which generates this asymmetry
of the energy density of the soliton.
  Thus, for example, the energy radiated by the soliton  comes mainly from the tail. When the velocity of the incoming soliton is close to its critical value the tail get badly deformed and the energy radiated from the tail is large. Below the critical velocity, {\it i.e.} when the soliton gets trapped inside the hole we have sometimes observed that the peak of the radiated energy is greater than the maximum of energy density of the soliton itself.
  In such cases one might even observe a false quantum-like reflection. This false reflection can be observed numerically unless one is careful and looks at the fields in some detail.
    While the soliton is inside the hole the magnitude of the back radiation may suggest
    its reflection. Only careful analysis of such a case would clarify the situation.
   In one case to be absolutely certain what has happened we had to make the hole very wide
   to be certain that the soliton remained trapped in it.
     This was the case, for example, when we looked at the $n=6$ model and the soliton
     was sent with velocity  $v=0.3$ towards a hole of depth -0.5. This soliton got trapped inside
     the hole and radiated away a very large amount of radiation. Figure 13 shows the energy density
     seen in this case when the radiated energy which has travelled backwards is higher than the main energy of soliton which is trapped inside the hole. One can also see from this figure that the front part
     of the soliton  has not shown any significan deformation while the tail part of the energy density has been deformed greatly and has given off almost all the radiated energy.

\begin{figure}
\begin{center}
\includegraphics[angle=270,width=4.5cm]{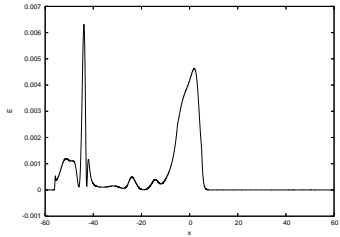}
\caption{Type-II models: Energy of a trapped soliton for $n$=6.}
\end{center}
\end{figure}

We have also observed that in the cases when the velocity is below its critical values the front part of the soliton including its centre of mass can exit the hole while the tail  still remains inside it. Then, because the tail has remained in the hole, the soliton gets pulled back into the hole and so remains trapped inside it.

\section{Conclusion}

 We have considered two classes of topological soliton models, presented by Bazeia et al \cite{one}
 which are generalisations of the Sine-Gordon model.
 Both classes depend on an integer parameter $n$ which,  for the type I models reduces to
 the Sine-Gordon model when $n=2$, while in the  type II case the Sine-Gordon case
 corresponds to $n$=1. Both classes of models have shown a behaviour which is similar to what was observed
 in the Sine-Gordon case in some recent research work, namely that the scattering by a hole is inelastic while the scattering by a barrier is nearly elastic. During the scattering on barriers the solitons
 behave very much like point-like particles. Their initial kinetic energy
 is converted into the potential energy needed to overcome the barrier and there is very little
 energy left to excite the internal degrees of freedom of the soliton.
 This has been confirmed by our estimates of the critical velocities for the transmission
 over the barrier which is based on this assumption. We have shown this to be true
 in all models ({\it i.e.} for all values of $n$).
 
 The scattering on the holes is different. When a soliton enters a hole it gains an extra
 energy which is then, in part, converted into the energy of internal oscillations.
 This is again seen for all the models we have looked at. Why this extra energy is used this way
 is not completely clear at the moment. We have tried to find an explanation in terms
 of equipartition of energy etc but have not managed to prove that this is really the case.
 Hence this explanation is still lacking and our results have shown that this is a general
 phenomenon as the excitation of the internal degrees of freedom takes place in the all models
 studied by us and is more or less similar in magnitude.
 
 We have also seen that the energy of solitons after the scattering is distributed differently into the vibrational and radiation energies. And so the soliton which converts most of its extra energy into
 the vibrational energy, rather than the radiation energy, is more able to be transmitted
over  the obstruction with the least velocity.

Solitons of the type-II models, except for the Sine-Gordon case ($n=1$), have asymmetrical field configurations and have exhibited what we have called the tail effect in their scattering on potential holes. The perturbation due to the potential hole has an uneven effect on the body of solitons. While it produces
 very little deformation  or the radiation in the front part of the solitons, their tails are greatly deformed and most of the radiation is produced there too. When the initial velocities of the solitons
 are close to their critical values, the radition energy, which is sent backwards, is very large and
 often more peaked than the energy density of the soliton itself. This
 could easily lead to some misunderstanding of what has happened; {\it i.e.}
 one may think that one has observed a quantum-like reflection while in reality
 this was only a trapping accompanied by some radiation sent backwards.

 Solitons, in some cases, behave like particles because of their localized structure. However, because of their extened structure they deviate from their particle behaviour in other cases. The symmetrical or asymmetrical field configurations play a great role on how the perturbations affect the solitons and the amount of radiation that is produced.

When a topological soliton (corresponding to $\tilde \lambda=1$) is placed inside a potential hole
it has an extra energy. This energy is then
transfered into its vibrational energy but some of it is radiated off.
 Moreover, when we looked at the energy of solitons scattered by a hole
we have found it to be almost the same as the energy of the solitons before the scattering although the final solitons were moving with somewhat smaller velocities. This implies that the scattering generated very little
radiation ({\it i.e.} also when the solitons were in the hole) and that most of the extra energy
was converted into the vibrational/excitation energy of the solitons and suprising little
was sent off as radiation.

We have also observed that for $n=3$ type I model, the critical velocity is the least for all the cases
we have looked at  when considering the  scattering of a soliton on a hole of -0.5 depth or on a barrier of 0.4 height.

\end{document}